\documentclass[zpreprint,zbstdefault]{./zeus_paper}
\usepackage[english]{babel}
\usepackage{verbatim}
\usepackage{psfrag}
\chardef\usc=95
\chardef\til=126
\catcode`\@=11 
\DeclareRobustCommand\xdotspace{\futurelet\@let@token\@xdotspace}
\def\@xdotspace{%
  \ifx\@let@token.\else
  \ifx\@let@token\bgroup.\else
  \ifx\@let@token\egroup.\else
  \ifx\@let@token\/.\else
  \ifx\@let@token\ .\else
  \ifx\@let@token~.\else
  \ifx\@let@token!.\else
  \ifx\@let@token,.\else
  \ifx\@let@token:.\else
  \ifx\@let@token;.\else
  \ifx\@let@token?.\else
  \ifx\@let@token/.\else
  \ifx\@let@token'.\else
  \ifx\@let@token).\else
  \ifx\@let@token-.\else
  \ifx\@let@token\@xobeysp.\else
  \ifx\@let@token\space.\else
  \ifx\@let@token\@sptoken.\else
   .\space
   \fi\fi\fi\fi\fi\fi\fi\fi\fi\fi\fi\fi\fi\fi\fi\fi\fi\fi}
\catcode`\@=12 

\newcommand{\stru}[2]{%
   \relax\ifmmode\hbox{\vrule height#1 depth#2 width0pt}%
   \else\vrule height#1 depth#2 width0pt\fi}

\newcommand{\Ronum}[1]{\uppercase\expandafter{\romannumeral#1}}
\newcommand{\ronum}[1]{\expandafter{\romannumeral#1}}
\DeclareRobustCommand{\LaTeXZ}{%
  \LaTeX\kern-.05em4\kern-.1em
  {\raisebox{-0.2ex}{$\scriptstyle\text{ZEUS}$}}\xspace}

\DeclareMathAlphabet{\mathbf}{OT1}{cmr}{bx}{sl}
\newcommand{\eVdist}{\kern-0.06667em}

\newcommand{\gev}{{\,\text{Ge}\eVdist\text{V\/}}}

\newcommand{\pbi}{\,\text{pb}^{-1}}

\newcommand{\Tesla}{\,\text{T}}

\newcommand{\slashfrac}[2]{%
  \raisebox{0.5ex}{\ensuremath #1}\kern-0.12em/\kern-0.08em
  \raisebox{-.8ex}{\ensuremath #2}}

\newcommand{\sqr}[3]{%
    {\vcenter{\hrule height.#3ex\hbox{\vrule width.#2ex height#1ex
     \kern#1ex\vrule width.#3ex}\hrule height.#2ex}}}

\catcode`\@=11 
\newcommand{\parenbar}{\mathpalette\p@renb@r}
\def\p@renb@r#1#2{\vbox{%
  \ifx#1\scriptscriptstyle \dimen@.7em\dimen@ii.2em\else
  \ifx#1\scriptstyle \dimen@.8em\dimen@ii.25em\else
  \dimen@1em\dimen@ii.4em\fi\fi \offinterlineskip
  \ialign{\hfill##\hfill\cr
    \vbox{\hrule width\dimen@ii}\cr
    \noalign{\vskip-.3ex}%
    \hbox to\dimen@{$\mathchar300\hfil\mathchar301$}\cr
    \noalign{\vskip-.3ex}%
    $#1#2$\cr}}}
\catcode`\@=12 

\newcommand{\rnge}{\hbox{$\,\text{--}\,$}}

\newcommand{\IP}{{\rm I$\kern-0.01667em$P}\xspace}

\mathchardef\qsm=63
\mathchardef\pls=43
\mathchardef\mns=512
\mathchardef\plm=518
\mathchardef\eql=61
\mathchardef\smallleft=300
\mathchardef\smallright=301
\mathchardef\les=316
\mathchardef\gre=318
\mathchardef\leq=532
\mathchardef\grq=533

\catcode`\@=11 
\newcounter{pict@width}
\newcounter{pict@height}
\newlength{\pict@scale}
\newcommand{\psfigadd}[4]{%
\setcounter{pict@width}{1*\ratio{#2+\pict@scale/2}{\pict@scale}}
\setcounter{pict@height}{1*\ratio{#3+\pict@scale/2}{\pict@scale}}
\setlength{\unitlength}{\pict@scale}
\hbox to #2{\hspace{-\fill}\begin{picture}(\thepict@width,\thepict@height)
\put(0,0){\psfig{figure=#1,width=#2,height=#3,clip=}}
\SetScale{0.283466457}
\SetWidth{1.763889}
{#4}
\end{picture}}
}
\newcounter{pict@widthfst}
\newcounter{pict@widthscd}
\newcounter{pict@widthtot}
\newcommand{\psfigaddtwo}[7]{%
\setcounter{pict@widthfst}{1*\ratio{#2+\pict@scale/2}{\pict@scale}}
\setcounter{pict@widthscd}{1*\ratio{#2+#4+\pict@scale/2}{\pict@scale}}
\setcounter{pict@widthtot}{1*\ratio{#2+#4+#6+\pict@scale/2}{\pict@scale}}
\setcounter{pict@height}{1*\ratio{#3+\pict@scale/2}{\pict@scale}}
\setlength{\unitlength}{\pict@scale}
\hbox{\hspace{-\fill}\begin{picture}(\thepict@widthtot,\thepict@height)
\put(0,0){\psfig{figure=#1,width=#2,height=#3,clip=}}
\put(\thepict@widthscd,0){\psfig{figure=#5,width=#6,height=#3,clip=}}
\SetScale{0.283466457}
\SetWidth{1.763889}
{#7}
\end{picture}}
}
\newcommand{\psfigror}[4]{%
\setcounter{pict@width}{1*\ratio{#2+\pict@scale/2}{\pict@scale}}
\setcounter{pict@height}{1*\ratio{#3+\pict@scale/2}{\pict@scale}}
\setlength{\unitlength}{\pict@scale}
\hbox{\begin{picture}(\thepict@width,\thepict@height)
\put(0,\thepict@height){\psfig{figure=#1,width=#3,height=#2,clip=,angle=270}}
\SetScale{0.283466457}
\SetWidth{1.763889}
{#4}
\end{picture}}
}
\newcommand{\psfigrol}[4]{%
\setcounter{pict@width}{1*\ratio{#2+\pict@scale/2}{\pict@scale}}
\setcounter{pict@height}{1*\ratio{#3+\pict@scale/2}{\pict@scale}}
\setlength{\unitlength}{\pict@scale}
\hbox{\begin{picture}(\thepict@width,\thepict@height)
\put(0,0){\psfig{figure=#1,width=#3,height=#2,clip=,angle=90}}
\SetScale{0.283466457}
\SetWidth{1.763889}
{#4}
\end{picture}}
}
\catcode`\@=12 
\newlength\listtextwidth

\catcode`\@=11 
\newlength{\@tabfninsert}
\newlength{\@tabfnwidth}
\newcommand{\tabfootnote}[2]{%
  \setlength{\@tabfninsert}{0.8em}
  \setlength{\@tabfnwidth}{\textwidth}
  \addtolength{\@tabfnwidth}{-\@tabfninsert}
  \addtolength{\@tabfnwidth}{-0.4em}
  \noindent\makebox[\@tabfninsert][r]{\footnotesize$^{#1}$\hfil}\hfill%
  \parbox[t]{\@tabfnwidth}{\footnotesize #2\hfill}}
\catcode`\@=12 

\def\citeCTD{{\cite{%
nim:a279:290,*npps:b32:181,*nim:a338:254%
}}\xspace}
\def\citeCAL{{\cite{%
nim:a309:77,*nim:a309:101,*nim:a321:356,*nim:a336:23%
}}\xspace}

\begin{document}

\prepnum{DESY--09--077}

\title{
Measurement of {\boldmath $J/\psi$} helicity distributions in inelastic
photoproduction at HERA
}

\author{ZEUS Collaboration}
\draftversion{}
\date{\today}

\abstract{
The $J/\psi$ decay angular distributions have been measured 
in inelastic photoproduction in $e p$ collisions with the ZEUS 
detector at 
HERA, using an integrated luminosity of 468 \( \pbi \).
The range in photon-proton centre-of-mass energy, $W$, was 
50 $< W <$ 180 \( \gev \). 
The $J/\psi$ mesons were identified through their decay into 
muon pairs. 
The polar and azimuthal angles of the $\mu^+$ were measured in 
the $J/\psi$ rest frame and compared to theoretical predictions 
at leading and next-to-leading order in QCD.
}

\makezeustitle

\begin{center}                                                                                     
{                      \Large  The ZEUS Collaboration              }                               
\end{center}                                                                                       
  S.~Chekanov,                                                                                     
  M.~Derrick,                                                                                      
  S.~Magill,                                                                                       
  B.~Musgrave,                                                                                     
  D.~Nicholass$^{   1}$,                                                                           
  \mbox{J.~Repond},                                                                                
  R.~Yoshida\\                                                                                     
 {\it Argonne National Laboratory, Argonne, Illinois 60439-4815, USA}~$^{n}$                       
\par \filbreak                                                                                     
  M.C.K.~Mattingly \\                                                                              
 {\it Andrews University, Berrien Springs, Michigan 49104-0380, USA}                               
\par \filbreak                                                                                     
  P.~Antonioli,                                                                                    
  G.~Bari,                                                                                         
  L.~Bellagamba,                                                                                   
  D.~Boscherini,                                                                                   
  A.~Bruni,                                                                                        
  G.~Bruni,                                                                                        
  F.~Cindolo,                                                                                      
  M.~Corradi,                                                                                      
\mbox{G.~Iacobucci},                                                                               
  A.~Margotti,                                                                                     
  R.~Nania,                                                                                        
  A.~Polini\\                                                                                      
  {\it INFN Bologna, Bologna, Italy}~$^{e}$
% ...
\newpage                                                                                           
\enlargethispage{5cm}                                                                              
$^{\    1}$ also affiliated with University College London,                                        
United Kingdom\\                                                                                   
% ...
$^{\    5}$ now at Institute of Aviation, Warsaw, Poland \\                                        
% ...
%
\newpage   
                                                           %
                                                           %
\begin{tabular}[h]{rp{14cm}}                                                                       
$^{a}$ &  supported by the Natural Sciences and Engineering Research Council of Canada (NSERC) \\  
$^{b}$ &  supported by the German Federal Ministry for Education and Research (BMBF), under        
          contract Nos. 05 HZ6PDA, 05 HZ6GUA, 05 HZ6VFA and 05 HZ4KHA\\                            
$^{c}$ &  supported in part by the MINERVA Gesellschaft f\"ur Forschung GmbH, the Israel Science   
          Foundation (grant No. 293/02-11.2) and the US-Israel Binational Science Foundation \\    
$^{d}$ &  supported by the Israel Science Foundation\\                                             
$^{e}$ &  supported by the Italian National Institute for Nuclear Physics (INFN) \\                
$^{f}$ &  supported by the Japanese Ministry of Education, Culture, Sports, Science and Technology 
          (MEXT) and its grants for Scientific Research\\                                          
$^{g}$ &  supported by the Korean Ministry of Education and Korea Science and Engineering          
          Foundation\\                                                                             
$^{h}$ &  supported by the Netherlands Foundation for Research on Matter (FOM)\\                   
$^{i}$ &  supported by the Polish State Committee for Scientific Research, project No.             
          DESY/256/2006 - 154/DES/2006/03\\                                                        
$^{j}$ &  partially supported by the German Federal Ministry for Education and Research (BMBF)\\   
$^{k}$ &  supported by RF Presidential grant N 1456.2008.2 for the leading                         
          scientific schools and by the Russian Ministry of Education and Science through its      
          grant for Scientific Research on High Energy Physics\\                                   
$^{l}$ &  supported by the Spanish Ministry of Education and Science through funds provided by     
          CICYT\\                                                                                  
$^{m}$ &  supported by the Science and Technology Facilities Council, UK\\                         
$^{n}$ &  supported by the US Department of Energy\\                                               
$^{o}$ &  supported by the US National Science Foundation. Any opinion,                            
findings and conclusions or recommendations expressed in this material                             
are those of the authors and do not necessarily reflect the views of the                           
National Science Foundation.\\                                                                     
$^{p}$ &  supported by the Polish Ministry of Science and Higher Education                         
as a scientific project (2009-2010)\\                                                              
$^{q}$ &  supported by FNRS and its associated funds (IISN and FRIA) and by an Inter-University    
          Attraction Poles Programme subsidised by the Belgian Federal Science Policy Office\\     
$^{r}$ &  supported by an FRGS grant from the Malaysian government\\                               
\end{tabular}                                                                                      
                                                           %
                                                           %


\pagenumbering{arabic} 
\pagestyle{plain}
\section{Introduction}
\label{sec:int}
In the HERA photoproduction regime, where the virtuality of the 
exchanged photon is small, the production of inelastic $J/\psi$ 
mesons is dominated by boson-gluon fusion: a photon emitted 
from the incoming lepton 
interacts with a gluon coming from the proton to produce a 
$c \bar{c}$ pair which subsequently forms a $J/\psi$ meson.
Production of $J/\psi$ through boson-gluon fusion can be calculated 
using perturbative 
Quantum Chromodynamics (pQCD) in the colour-singlet (CS), 
in the non-relativistic QCD (NRQCD)
~\cite{ppnp:47:141,hep-ph-0412158} and in the $k_T$-factorisation 
frameworks~\cite{pl:b428:377,epj:c27:87}. 

In the CS approach, only the colourless $c \bar{c}$ pair produced in the hard 
subprocess can lead to a physical $J/\psi$ state.
In the NRQCD approach, a $c \bar{c}$ pair emerging from the hard process 
in a colour-octet (CO) state can also evolve into a $J/\psi$ state 
with a probability proportional to universal long-distance matrix elements 
(LDME) that are obtained from experiment. 
In the $k_T$-factorisation approach, the effects of non-zero incoming parton 
transverse momentum are taken into account. 
Cross sections are then calculated in the CS approach as a 
convolution of unintegrated (transverse-momentum dependent) 
parton densities and leading-order (LO) off-shell matrix elements.

At HERA, measurements of inelastic $J/\psi$ differential cross 
sections~\cite{epj:c27:173,epj:c25:25} are   
reproduced by a next-to-leading-order (NLO) QCD 
calculation~\cite{pl:b348:657,np:b459:3} performed 
in the CS framework. The measurements are also reasonably
well described by LO CS plus CO 
calculations~\cite{prl:76:4128,epj:c6:493,pr:d62:34004}, 
with LDME as determined in a LO analysis of hadroproduction and $B$-decay 
data~\cite{pr:d55:5269,pr:d54:2005,pr:d54:4312,pr:d55:4098}.

The polar and azimuthal distributions of the $J/\psi$ decay leptons 
in the $J/\psi$ rest frame may be used to distinguish between CS and CO 
models.
These helicity distributions are 
expected to be different as a function of the $J/\psi$ transverse momentum,   
$p_T$, and inelasticity, the fraction of the incident photon energy 
carried by the $J/\psi$ in the proton rest frame, $z$~\cite{pr:d57:4258}.

Helicity-distribution measurements have already been performed by the
ZEUS~\cite{epj:c27:173} and H1~\cite{epj:c25:25}
collaborations. 
In the final study presented here, $J/\psi$ mesons were identified using the 
decay mode $J/\psi \rightarrow \mu^{+} \mu^{-}$ and were measured in the range 
$50 < W < 180$ GeV, where $W$ is the photon-proton centre-of-mass energy.
The data sample under study includes the data used in the 
previously published ZEUS analysis~\cite{epj:c27:173} and corresponds to 
an increase in statistics of a factor of 12.

\section{Experimental set-up}
\label{sec:setup}

The analysis presented here is based on data collected by the ZEUS detector 
at HERA in the period 1996--2007. 
In 1998--2007 (1996--1997), HERA provided electron\footnote{Here and 
in the following, 
the term ``electron'' denotes 
generically both the electron ($e^-$) and the positron ($e^+$).}
beams of energy $E_e$ = 27.5 GeV and proton beams of energy $E_p = 920 
~(820)$ GeV, resulting in a centre-of-mass energy of $\sqrt{s} = 318 ~(300)$ 
GeV, corresponding to an integrated luminosity of $430 \pm 11 ~(38 \pm 0.6)$
pb$^{-1}$.  
 
A detailed description of the ZEUS detector can be found
elsewhere~\cite{pl:b293:465,zeus:1993:bluebook}. A brief outline of the 
components that are most relevant for this analysis is given below. 
Charged particles were tracked in the central tracking detector 
(CTD)~\citeCTD, which operated in a magnetic field of 
$1.43\Tesla$ provided by a thin superconducting coil. 
Before the 2003--2007 running period, the ZEUS tracking system was 
upgraded with a silicon microvertex detector (MVD)~\cite{nim:a581:656}. 
The high-resolution uranium--scintillator calorimeter (CAL)~\citeCAL 
consisted of three parts: the forward (FCAL), the barrel (BCAL) and the 
rear (RCAL) calorimeters\footnote{The ZEUS coordinate system is a 
right-handed Cartesian system, with the $Z$ axis pointing in the proton beam 
direction, referred to as the ``forward direction'', and the $X$ axis pointing 
towards the centre of HERA.
The coordinate origin is at the nominal interaction point. 
The polar angle, $\theta$, is measured with respect to the proton beam 
direction. The pseudorapidity is defined as
$\eta$=--ln(tan $\frac{\theta}{2}$).}. 

Muons were identified by tracks in the barrel and rear 
muon chambers (BMUON and RMUON)~\cite{nim:a333:342}.
The muon chambers
were placed inside and outside the magnetised iron yoke surrounding the CAL.
The barrel and rear inner muon chambers (BMUI and RMUI) covered the 
polar-angle regions  $34^\circ < \theta < 
 135^\circ$ and $135^\circ < \theta < 171^\circ$, respectively.
The luminosity was measured using the Bethe--Heitler reaction 
$ep \rightarrow e \gamma p$ with the luminosity detector 
which consisted of a lead--scintillator calorimeter 
\cite{desy-92-066,*zfp:c63:391,*acpp:b32:2025} and, after 2002, 
an additional magnetic spectrometer\cite{nim:a565:572} system. 

\section{Event selection}
\label{sec:event}

Inelastic events are often selected using the inelasticity, $z$. In this 
analysis, however, the events were selected using the transverse 
momentum, $p_T$, of the $J/\psi$ and additional activity in the detector. 
This kind of selection permits direct comparisons with 
the different theoretical predictions~\cite{pr:d57:4258}.

The online and offline selections as well as the reconstruction of the 
kinematic variables closely follow a previous 
analysis~\cite{epj:c27:173}.

Online, the BMUI and RMUI chambers were used to tag muons by matching 
segments in the muon chambers with tracks in the CTD/MVD, as well as 
with energy deposits in the CAL consistent with the passage of a 
minimum-ionising particle (m.i.p.).
 
Offline, an event was accepted if it had two tracks forming a $J/\psi$
candidate. 
One track had to be identified in the inner muon chambers and 
matched to a m.i.p. cluster in the CAL. It was required to have a momentum 
greater than 1.8 GeV if it was in the rear region, or a transverse 
momentum greater than 1.4 GeV if in the barrel region. The other track 
had to be matched to a m.i.p. cluster in the CAL and was required 
to have a transverse momentum greater than 0.9 GeV. 
Both tracks were restricted to the 
pseudorapidity region $|\eta |<$ 1.75.
To reject cosmic rays, events in which the angle between the two muon 
tracks was larger than 174$^{\circ}$ were removed. 

The $p_T$ of the $J/\psi$ candidate was required to be larger than 1 GeV. 
In addition, events were required to have 
an energy deposit larger than 1 GeV in a cone of 35$^{\circ}$ around the 
forward direction (excluding possible calorimeter deposits 
due to the decay muons). 
According to Monte Carlo (MC) simulations, these requirements completely 
reject exclusively produced $J/\psi$ mesons ($e p \rightarrow e p J/\psi$) 
as well as 
proton-diffractive events ($e p \rightarrow e Y J/\psi$) in which 
the mass of 
the proton dissociative state, $M_Y$, is below 4.4 GeV. 
To further reduce diffractive background, events were also required 
to have at least one additional track with a transverse momentum larger 
than 0.125 GeV and  pseudorapidity $|\eta |<$ 1.75.

\section{Kinematic variables and signal extraction} 
\label{sec:kinematic}

The photon-proton centre-of-mass energy, $W$, is:
\begin{equation}
\label{eq:W_def}
W^2 = (P +q)^2 ,
\end{equation}
where $P$ and $q$ are the four--momenta of the incoming 
proton and exchanged photon, respectively. 
It was calculated using:   
\begin{equation}
\label{eq:W_comp}
W^2 = 2 E_p (E-p_Z) 
\end{equation}
where $(E-p_Z)$  is summed over all final-state energy-flow
objects~\cite{epj:c1:81,*thesis:briskin:1998}
(EFOs) which combine the information from calorimetry and tracking.

The inelasticity $z=  \frac{P \cdot p_{J/\psi} }{P \cdot q}$ 
was determined as:  
\begin{equation}
\label{eq:z_def}
z = \frac{(E-p_Z)_{J/\psi}}{(E-p_Z)},
\end{equation}
where $p_{J/\psi}$ is the four-momentum of the $J/\psi$ and 
$(E-p_Z)_{J/\psi}$ was calculated using 
the tracks forming the $J/\psi$. 

The kinematic region considered was $50<W< 180$~GeV where the acceptance 
was always above 10$\%$. 
A requirement of $E-p_Z < 20$~GeV restricted the virtuality of the 
exchanged photon $Q^2 = -q^2 \lesssim 1$~GeV$^2$, with a 
median of $\approx 10^{-4}$~GeV$^2$. The elimination of deep inelastic 
scattering events was independently confirmed by searching for scattered 
electrons in the CAL~\cite{nim:a365:508}; none was found.

\begin{figure}[hbpt!]
\begin{center}
\unitlength1in
\psfig{figure=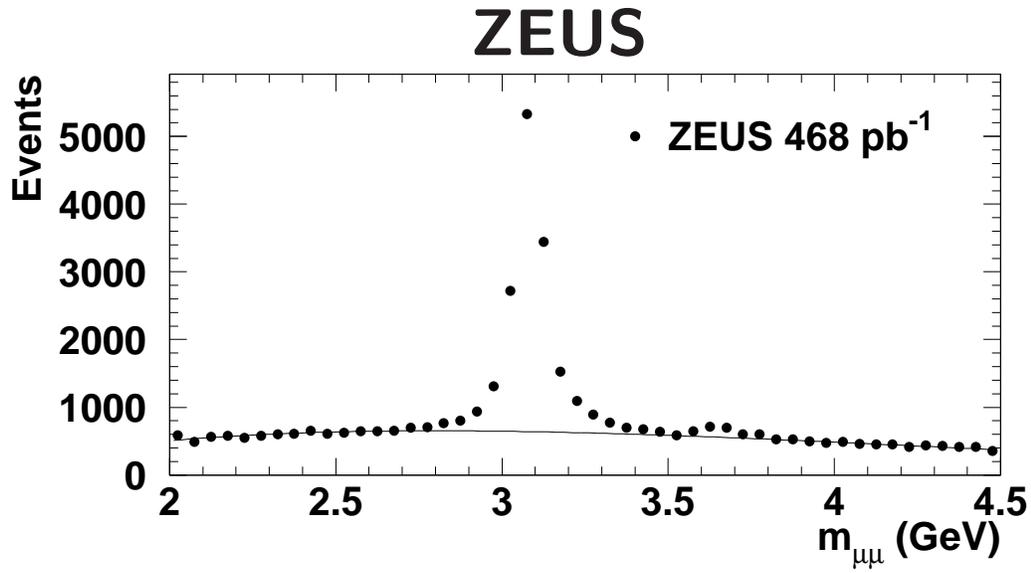,height=6.0in}
\put(-3.6,5.5){\Huge \bf \textsf{ZEUS}}
\vspace{-6.0cm}
\end{center}
\caption{Dimuon invariant mass, m$_{\mu \mu}$, spectrum in the phase-space 
region $50 < W < 180$ GeV, $z > 0.1$ and $p_T >$ 1 GeV. 
The continuous line represents the fitted background.}
\label{fig:mmumu}
\end{figure}

The invariant-mass spectrum of the $J/\psi$ candidates with 
$p_T >$ 1 GeV and  $z >$ 0.1 is shown in 
Fig.~\ref{fig:mmumu}. Both the $J/\psi$ and $\psi^{'}$ peaks are visible. 
The background was estimated by fitting the product of a second-order 
polynomial and an exponential to the region outside the invariant-mass 
window, 2.85--3.3 GeV.
The number of $J/\psi$ events was obtained by subtracting the number of 
background events estimated from the fit procedure from the total number
of events inside the invariant-mass window; 12310 $\pm$ 140 $J/\psi$ events 
were found.
As the signal to background ratio is large, 
the extracted number of $J/\psi$ events has
little sensitivity to the analytical form of the function used for the
background fit.

\section{Monte Carlo and background evaluation}
\label{sec:montecarlo}

The inelastic production of $J/\psi$ mesons 
was simulated 
using the {\sc Herwig} 5.8~\cite{cpc:67:465} 
MC generator, which generates events according to the 
LO diagrams of the boson-gluon-fusion process, $\gamma g \rightarrow 
J/\psi g$, as calculated in the framework of the CS model. 
This process is called a direct photon process, because 
the incoming photon couples to the $c$ quark directly.
The {\sc Herwig} MC sample was reweighted in $p_T$, $z$ and $W$ in order 
to give the best description of the data.

There are other sources of $J/\psi$ mesons which were classified as background 
in the present analysis and were estimated either from MC models or 
previous measurements. Although the relative rate of each process is 
given below, the helicity distributions of these $J/\psi$ sources are 
poorly known, so the contributions were not subtracted.

Diffractive production of $J/\psi$ mesons 
with proton dissociation was simulated 
with the {\sc Epsoft}~\cite{thesis:kasprzak:1994} MC generator, which was 
tuned to describe such processes at HERA~\cite{thesis:adamczyk:1999}. 
This background is suppressed by the requirement on the tracks and by the 
cut on the minimum $p_T$ of the $J/\psi$. The overall contribution of this 
background is 6\%; it is largest in the lowest $p_T$ bin ($ 1~ \leq ~p_T~ 
\leq ~1.4$  GeV), where it is 7.5\%, and in the highest $z$ 
bin ($0.9~ \leq ~z~ \leq ~1$), where it is 66\%.

The production of  $J/\psi$ mesons originating from $B$-meson decays 
was simulated using the {\sc Pythia~6.2} MC generator~\cite
{cpc:135:238,*epj:c17:137,*hep-ph-0108264}. 
The beauty-quark mass was set to 4.75 GeV and the $B$ to $J/\psi$ 
branching ratio was set to the PDG value~\cite{pl:b667:1}.
According to the MC, 1.6\% of the observed $J/\psi$ events were from 
$B$-meson decays; the fraction is largest in the highest $p_T$ bin 
($ 4.2~ \leq ~p_T~ \leq ~10$  GeV), where it is  equal to 6.3\%, and in 
the lowest $z$ bin ($0.1~ \leq ~z~ \leq ~0.4 $), where it is 8.4\%.

The background from $\psi^{'}$ to $J/\psi$ decays
is expected to be around 15\%, as obtained using the 
direct measurement of the $\psi^{'}$ to $J/\psi$ cross section 
ratio~\cite{epj:c27:173} and the branching ratio of the $\psi'$ to $J/\psi$.
  
All generated events were passed through a full simulation of the ZEUS 
detector based on {\sc Geant} 3~\cite{tech:cern-dd-ee-84-1}. 
They were then subjected to the same 
trigger requirements and processed by the same reconstruction program 
as the data.

\section{Reconstruction of the helicity parameters}
\label{sec:param}

The helicity analysis was performed in the so-called ``target 
frame''~\cite{pr:d57:4258}, i.e. the $J/\psi$ rest frame with the 
axes $Z^{\prime}= -Z$ and $Y^{\prime}$ along the vector 
$\vec{q^{\prime}} \times (- \vec{P^{\prime}})$, where 
$\vec{q^{\prime}}$ and $\vec{P^{\prime}}$ are the three-vectors associated 
with the exchanged photon and incoming proton. 
The polar and azimuthal angles of the $\mu^{+}$ in this frame are denoted 
$\theta^{\star}$ and $\phi^{\star}$. 

The differential cross sections in $\theta^{\star}$ and 
$\phi^{\star}$ can be parametrised as~\cite{pr:d57:4258}:
\begin{equation}
\frac{d \sigma}{d \cos \theta^{\star}} \propto
1 + \lambda  \cos^2 \theta^{\star},
\label{eq:lambda}
\end{equation}
and
\begin{equation}
\frac{d \sigma}{d \phi^{\star}} \propto
1 + \frac{\lambda}{3} +
   \frac{\nu}{3} \cos 2 \phi^{\star},
\label{eq:nu}
\end{equation}
where $\lambda$ and $\nu$, the polar and azimuthal angular parameters, are
functions of $p_T$ and $z$.
The predictions for $\lambda$ and $\nu$ depend on the production 
mechanism. The value $\lambda = +1$ corresponds to $J/\psi$ mesons
fully transversally polarised, while $\lambda = -1$ corresponds to $J/\psi$
mesons fully longitudinally polarised.

The $\lambda$ and $\nu$ parameters were determined in bins of $z$ and $p_T$, 
each time integrating over the other variable.
As a function of $p_T$, the integration range for 
$z$ was set to $0.4 < z < 1$, thereby avoiding the region 
$0.1 < z < 0.4$ where the ratio of signal to combinatorial background is 
rather poor (0.52).
The integration range in $p_T$ started at $p_T$ = 1 GeV.

In the estimation of the parameters $\lambda$ and $\nu$, the 
helicity distributions of the background events present 
under the $J/\psi$ 
peak were added to the MC distributions. The shape of the background helicity 
distributions was taken from the side bands, while the number 
of background events was taken from the fits described in 
Section~\ref{sec:kinematic}.

The {\sc Herwig} MC generator-level distributions $dN/d \cos \theta^{\star}$ 
($dN/d \phi^{\star}$)
were re-weighted according to Eq. \ref{eq:lambda} (\ref{eq:nu}) within
a search grid of $\lambda$ ($\nu$) values. For each re-weighted distribution,
the value of $\chi^2$ was calculated from a comparison to the data.
The $\lambda$ ($\nu$) value providing the minimum $\chi^2$, 
$\chi^2_{\rm min}$, 
was taken as the central value. The parameter values with 
$\chi^2 = \chi^2_{\rm min} + 1$ were used to calculate the statistical 
uncertainties.
The $\chi^2_{\rm min}$ per degree of freedom were typically around one.
Equation~\ref{eq:lambda} was first used to extract $\lambda$, 
and then $\lambda$ was inserted into Eq. \ref{eq:nu} to extract $\nu$,
see Tables~\ref{tab:lanuvspt} and \ref{tab:lanuvsz}. 

\section{Systematic uncertainties}

The following sources of systematic uncertainties were investigated
(their effects are given in parentheses):

\begin{itemize}

\item muon chamber efficiencies: the BMUI and RMUI muon chamber efficiencies 
were extracted from the data using muon pairs coming from elastic 
$J/\psi$ events and from the process $\gamma \gamma \rightarrow \mu^+ \mu^-$.
These efficiencies are known up to an uncertainty of about $\pm 5 \%$
($< 5 \%$ of the statistical error); 
 
\item analysis cuts: this class comprises the systematic uncertainties due 
to the uncertainties in the measurement of momentum, transverse momentum and 
pseudorapidity of the muon decay tracks. Each cut was varied within a range 
determined by the resolution in the appropriate variable 
($< 5 \%$ of the statistical error); 

\item CAL energy scale: the CAL energies were varied by $\pm 5 \%$ in the
simulation, in accordance with the uncertainty in the CAL energy scale
(on average  $10 \%$ of the statistical error);
 
\item hadronic energy resolution: the $W$ and $z$ resolutions are dominated 
by the hadronic energy resolution affecting the quantity $(E-p_Z)$. The 
$(E-p_Z)$ resolution in the MC was smeared event by event by $\pm 20\%$ 
(on average $ 10 \%$ of the statistical error);

\item $p_T$, $W$ and $z$ spectra: the $p_T$, $W$  and $z$  spectra 
of the $J/\psi$ mesons 
in the {\sc Herwig} MC simulation were varied within ranges allowed by the 
comparison between data and simulation 
(on average $ 15 \%$ of the statistical error); 

\item additional track requirement: the kinematic cuts for the additional
track requirement were tightened and loosened in both data and MC 
(on average $15 \%$ of the statistical error); 

\item influence of diffractive contamination at high $z$ on the $\lambda$ 
and $\nu$ extractions as a function of $p_T$:
$\lambda$ and $\nu$ were extracted changing the $z$ integration range from 
$0.4 < z < 1$ to $0.4 < z < 0.9$ (on average $ 25 \%$ of the 
statistical error);

\item angular coverage in $\theta^{\star}$: $\cos \theta^{\star}$ 
was restricted to the range 
$-0.8 < \cos \theta^{\star} < 0.8$ in order to avoid low-acceptance regions
(on average $ 30 \%$ of the statistical error);    

\item invariant-mass window: the $J/\psi$ invariant mass window was 
enlarged by 50 MeV and tightened by 100 MeV 
(on average $30\%$ of the statistical error).

\end{itemize}
All of the above individual sources of systematic uncertainty were added 
in quadrature. 
No systematic uncertainties are quoted for $J/\psi$ coming from $B$-meson
decays and $J/\psi$ coming from $\psi^{'}$ decays.
The uncertainties on the integrated-luminosity determination and on 
the $J/\psi \rightarrow \mu^+\mu^-$ branching ratio, which would 
result in an overall shift of a cross section measurement, 
do not contribute to the measurement of the helicity parameters. 

\section{Results}
\label{sec:res}

\begin{figure}[hbpt!]
\unitlength1cm
\includegraphics[width=0.5\textwidth]{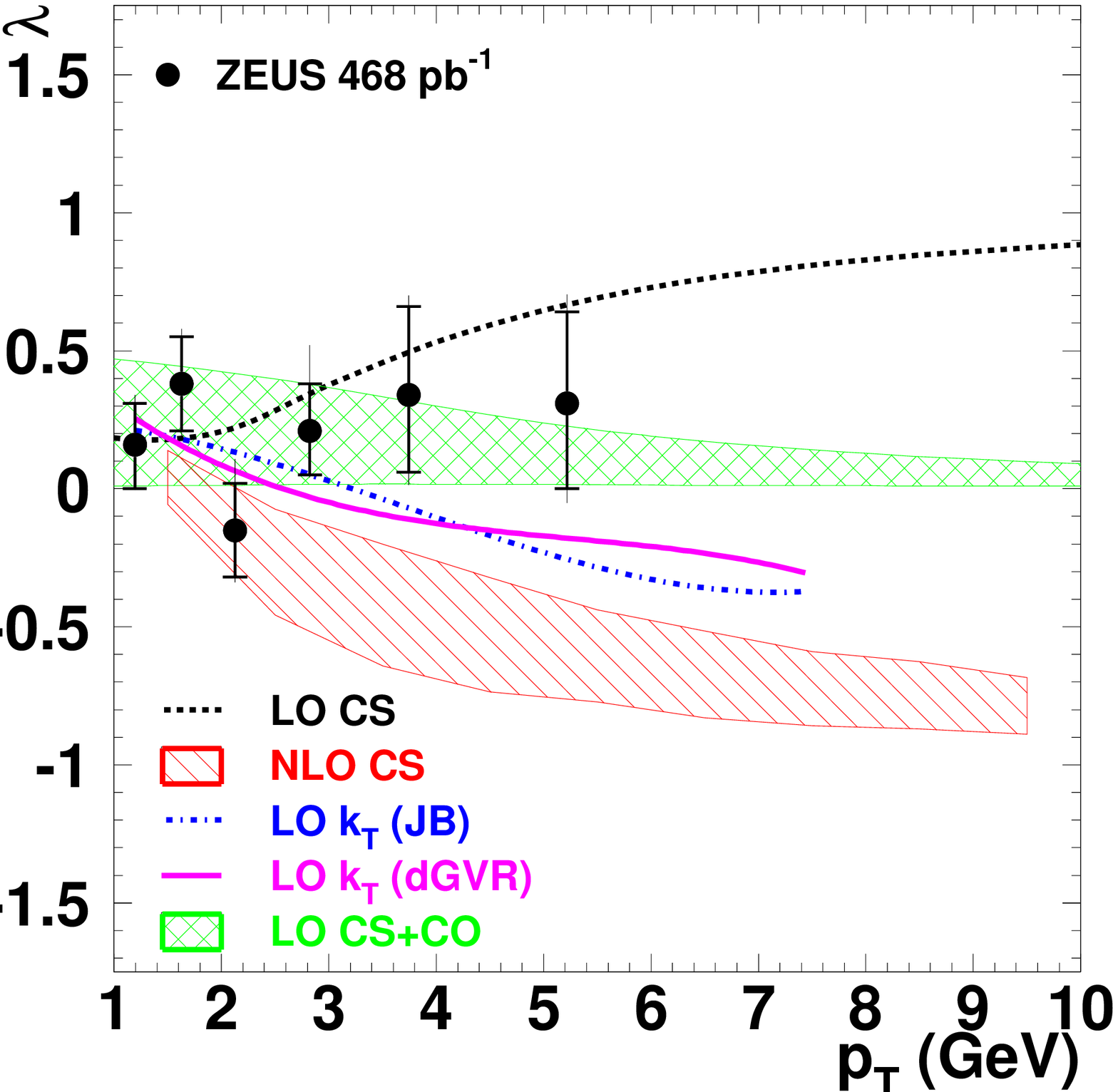}
\put(-1.1,7.5){\Huge \bf \textsf{ZEUS}}
\put(-2.0,1.2){\bf (a)}
\includegraphics[width=0.5\textwidth]{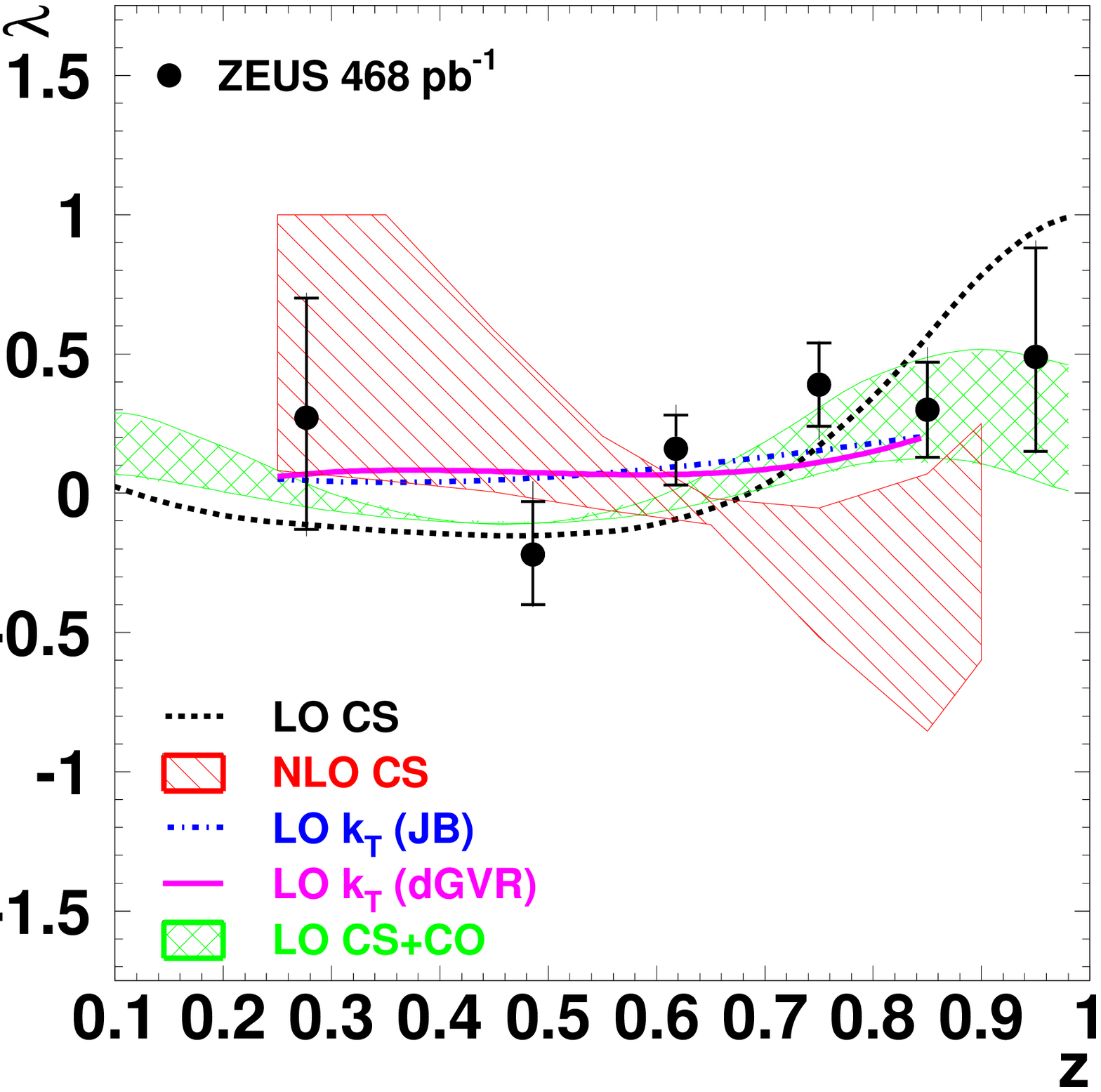}
\put(-2.0,1.2){\bf (b)}
\caption{The helicity parameter $\lambda$, measured in the target frame, 
as a function of (a) $p_T$, and (b) $z$.
The measurement is performed in the kinematic range 
$50 < W < 180 $ GeV, $0.1 < z < 1$ and $p_T > $ 1 GeV. 
The measurement as a function of 
$p_T$ is restricted to the kinematic range $0.4 < z < 1$.
The inner (outer) error bars correspond to the statistical (total) 
uncertainty.
The theoretical curves are described in the text.
}
\label{fig:lambda}
\end{figure}

\begin{figure}[hbpt!]
\unitlength1cm
\includegraphics[width=0.5\textwidth]{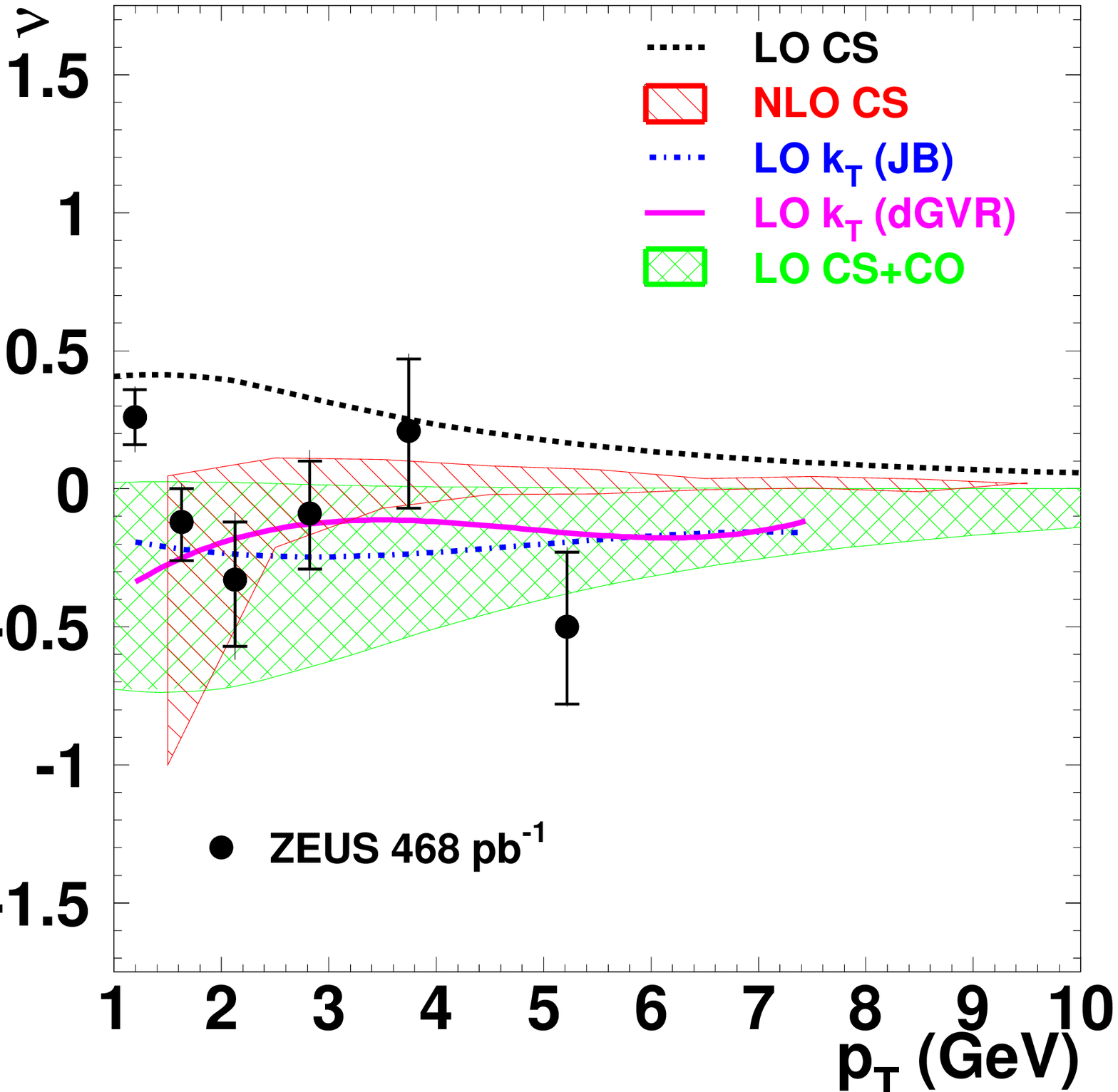}
\put(-1.1,7.5){\Huge \bf \textsf{ZEUS}}
\put(-2.0,1.2){\bf (a)}
\includegraphics[width=0.5\textwidth]{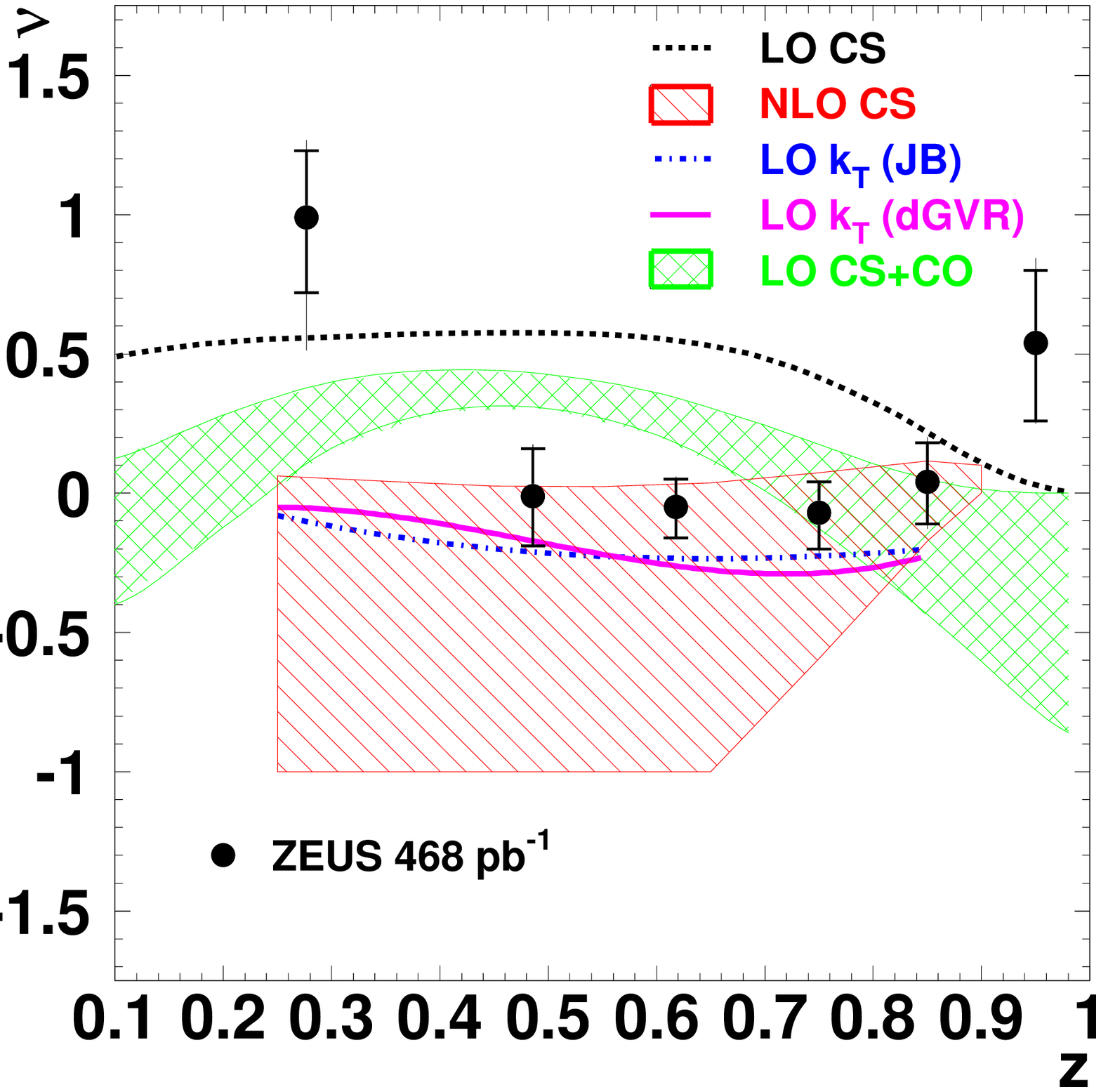}
\put(-2.0,1.2){\bf (b)}
\caption{The helicity parameter $\nu$, measured in the target frame, 
as a function of (a) $p_T$, and (b) $z$.
The measurement is performed in the kinematic range 
$50 < W < 180 $ GeV, $0.1 < z < 1$ and $p_T > $ 1 GeV. 
The measurement as a function of 
$p_T$ is restricted to the kinematic range $0.4 < z < 1$.
The inner (outer) error bars correspond to the statistical (total) 
uncertainty.
The theoretical curves are described in the text.
}
\label{fig:nu}
\end{figure}

The values of the parameter  
$\lambda$ are shown as a function of $p_T$ and $z$ in 
Fig.~\ref{fig:lambda}a) and b), respectively;
the values of the parameter  
$\nu$ are displayed in Fig.~\ref{fig:nu}a) and b). 
All the values are also listed in Tables~\ref{tab:lanuvspt} and 
\ref{tab:lanuvsz}.
The data indicate that the parameter $\lambda$ depends, if at all, only 
weakly on $p_T$ and rises slowly with $z$. 
The parameter $\nu$ does not seem to depend on  $p_T$, while it seems 
to increase at low and high $z$.

The data are compared to various theoretical predictions for 
photoproduction at $Q^2 = 0$. These 
predictions do not consider the polarisation due to 
$J/\psi$ coming from $\psi^{'}$ decays, 
$B$-meson decays and from diffractive processes. 
The curves identified by the label LO CS show
the LO prediction in the CS framework including both direct and 
resolved\footnote{In these resolved processes, the incoming photon 
does not couple to the $c$ quark directly, but via its hadronic component.
They are expected to contribute mainly to the region of $z <$ 0.4.}
processes.
The two lines identified by the label LO + $k_T$ (JB) and
LO + $k_T$ (dGRV), 
represent the predictions of a $k_T$-factorisation model
~\cite{jetp:88:471} using two different unintegrated gluon distributions
and including only direct processes. 
The band identified by the label NLO CS 
represents the predictions of a NLO 
calculation~\cite{arXiv:0901.4352,priv:artoisenet:2009} including only 
direct processes. 
The width of the band gives the uncertainties of the 
calculation due to variations of the 
renormalisation and factorisation scales. 
It stops at $z=0.9$ because no reliable 
predictions can be obtained near $z=1$ for this fixed-order calculation.
The band, identified by the label LO CS+CO, 
shows the LO prediction~\cite{pr:d57:4258} including both CS and CO 
terms, including both direct and resolved processes.
The width of the band results from the uncertainties in the values of the 
long-distance matrix elements.
In Fig.~\ref{fig:lambda}a) and \ref{fig:nu}a),   
with $z$ integrated up to $z = 1$, the LO CS+CO  
cross section is CO dominated.

None of the models provides predictions for both $\lambda$ and $\nu$
that agree well with the data everywhere in $z$ and $p_T$.
For $\lambda$ as a function of $p_T$, all the models roughly describe 
the data, with the NLO CS prediction providing the poorest
description. For high values of $p_T$, the polarisation in the data 
remains small, while LO CS predicts a progressive increase and NLO
CS and LO $k_T$ a progressive decrease.
The LO CS + CO prediction remains flat, with a small and 
positive value of $\lambda$.
For $\lambda$ as a function of $z$, all theoretical predictions are in 
rough agreement with the data.
The $p_T$ and $z$ dependencies of $\nu$ are not described by the LO CS
predictions, while the other models 
provide better descriptions of the data. 

\begin{figure}[hbpt!]
\unitlength1cm
\begin{center}
\includegraphics[width=0.50\textwidth]{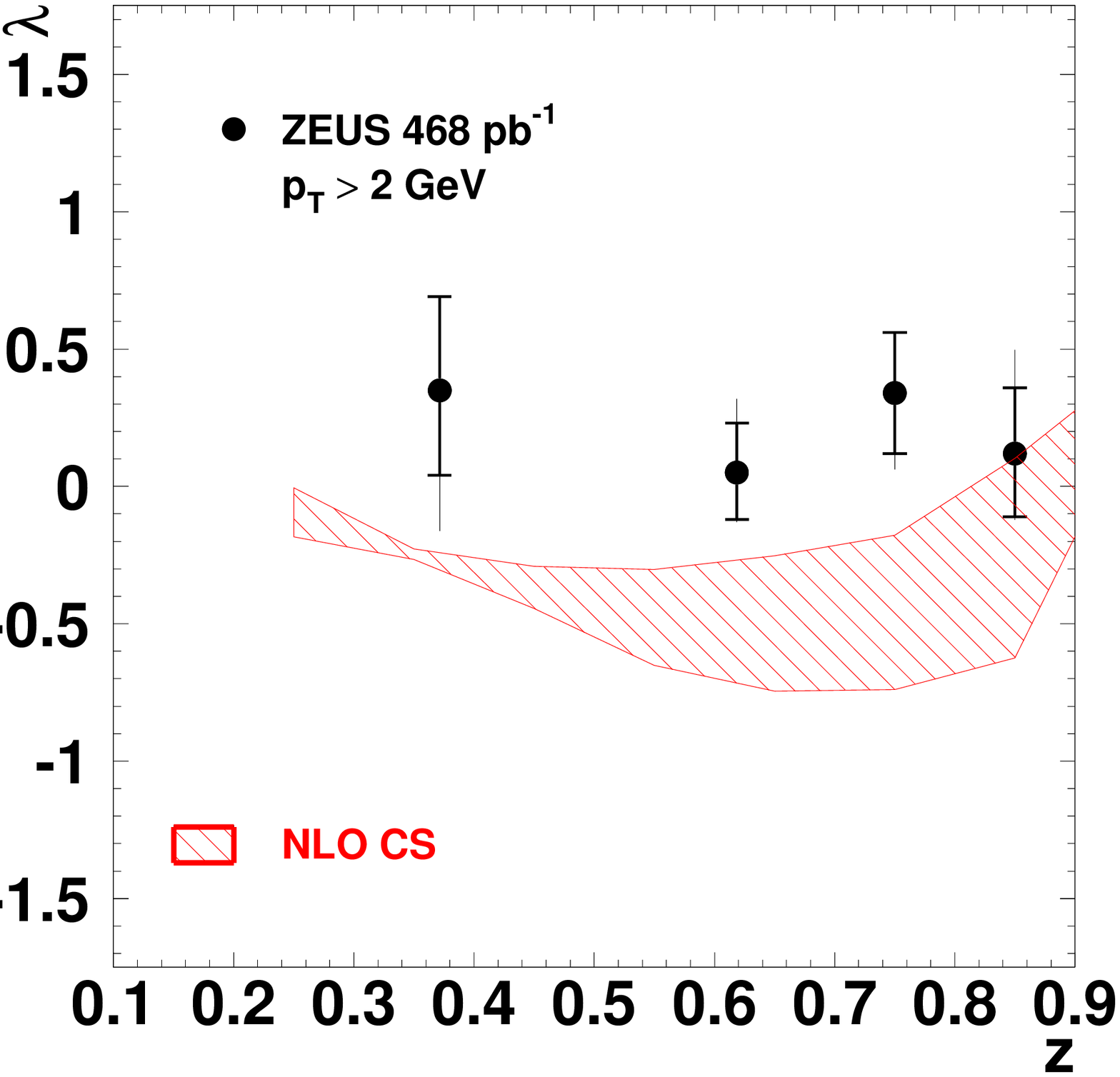}
\put(-1.1,7.5){\Huge \bf \textsf{ZEUS}}
\put(-2.0,1.2){\bf (a)}
\includegraphics[width=0.50\textwidth]{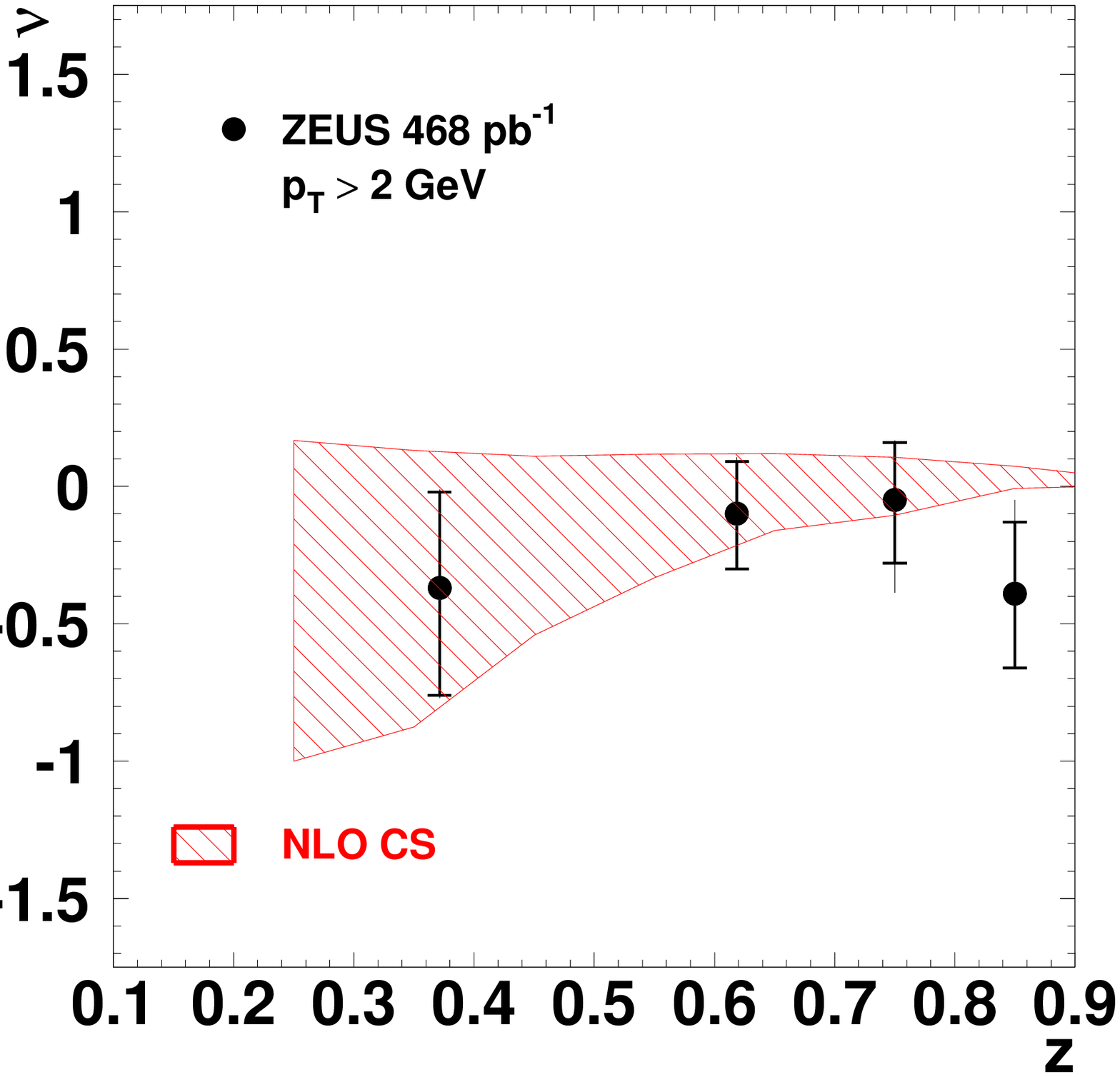}
\put(-2.0,1.2){\bf (b)}
\hfill
\includegraphics[width=0.50\textwidth]{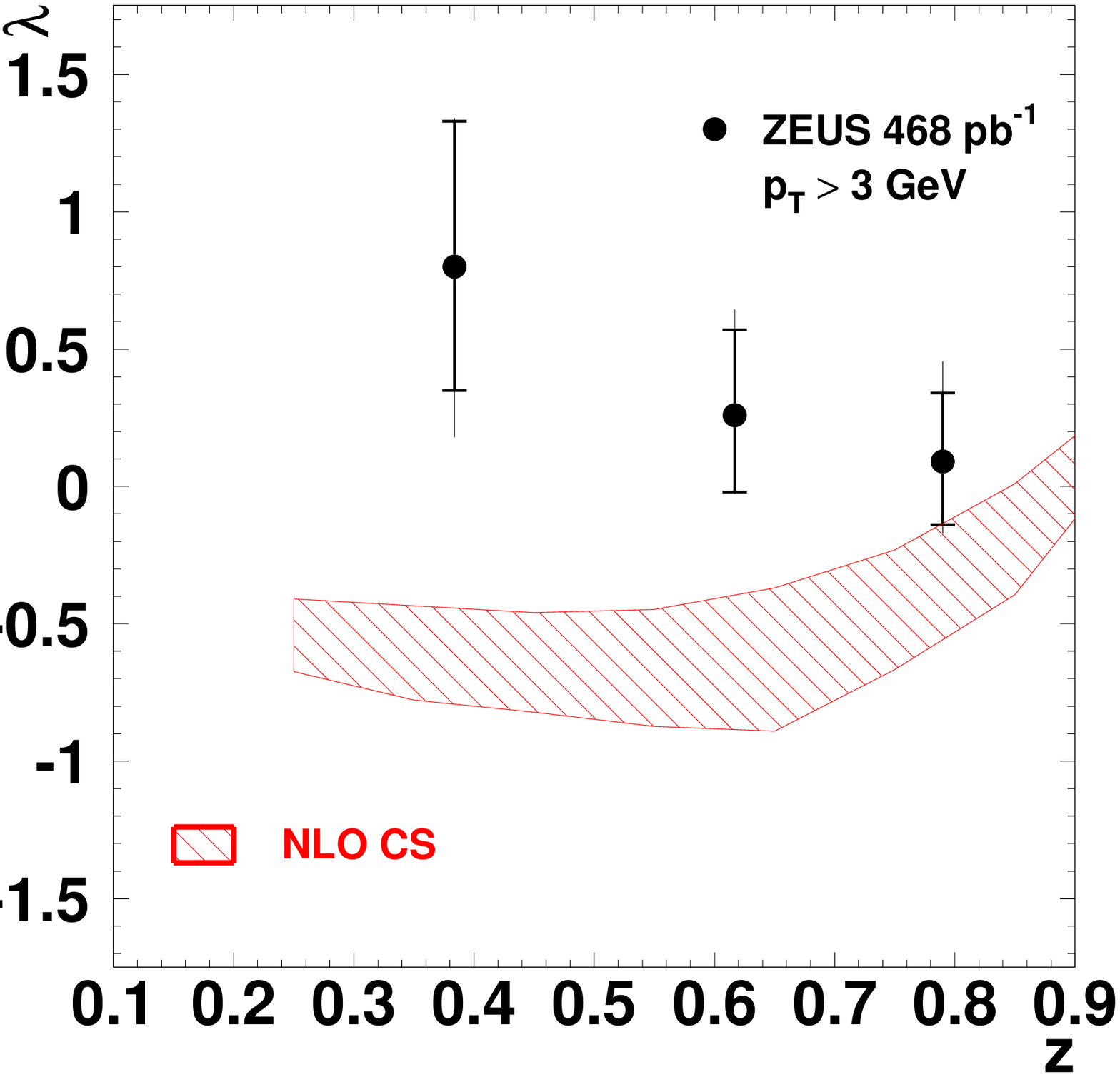}
\put(-2.0,1.2){\bf (c)}
\includegraphics[width=0.50\textwidth]{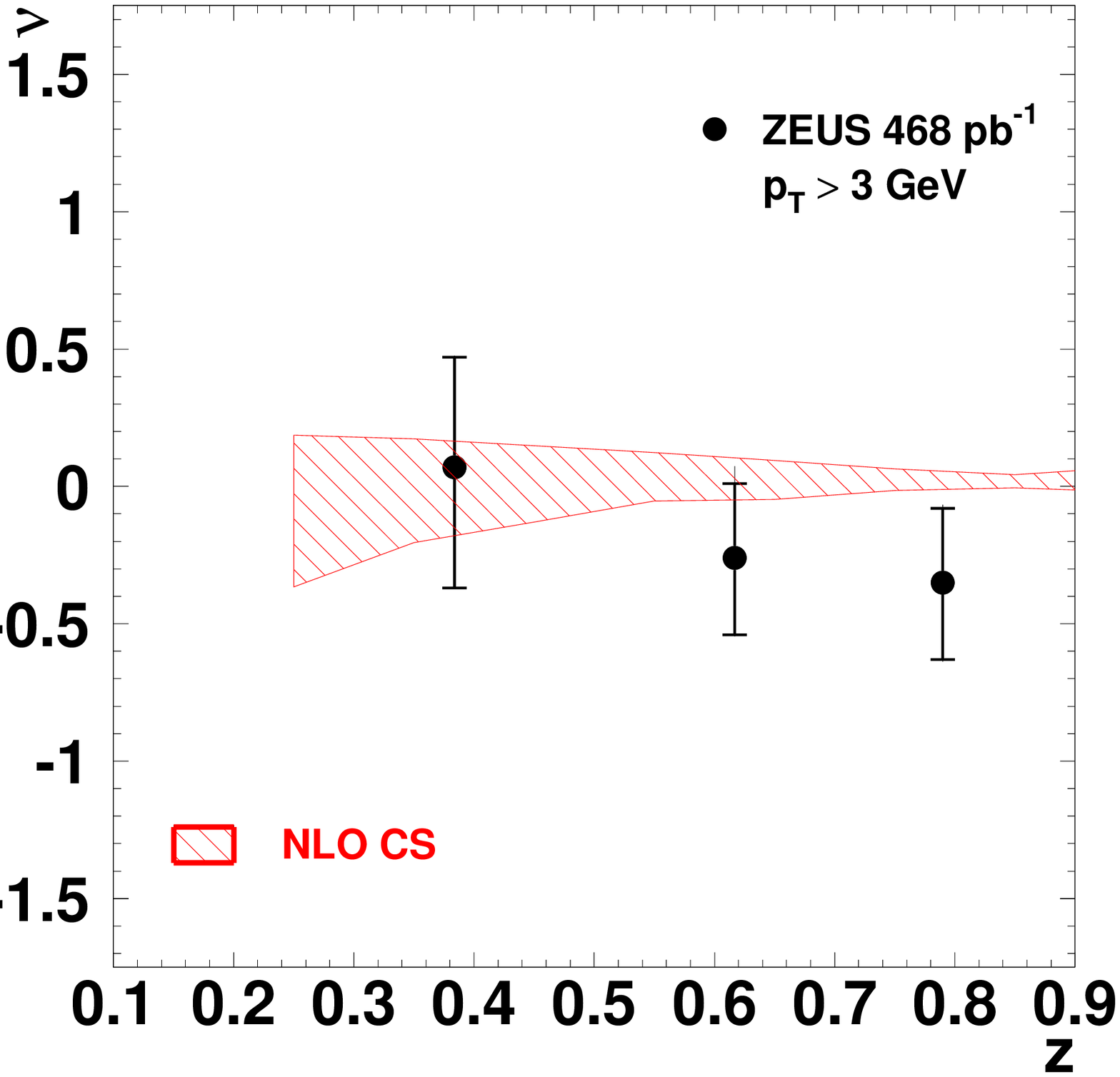}
\put(-2.0,1.2){\bf (d)}
\end{center}
\caption{Distributions of the helicity parameters (a), (c) $\lambda$ 
and (b), (d) $\nu$ as a function of $z$, measured in the target 
frame, for $50 < W < 180 $ GeV, $0.1 < z < 0.9$ and (a), (b) 
$p_T > $ 2 GeV and (c), (d) $p_T > $ 3 GeV.
The inner (outer) error bars correspond to the statistical (total) 
uncertainty. The theoretical bands are described in the text.
}
\label{fig:highpt}
\end{figure}

The NLO CS calculation for $p_T >$ 1 GeV suffers from large 
scale uncertainties connected to the presence of negative 
values of the diagonal components of the spin density matrix at 
$p_T \lesssim $ 1 GeV~\cite{arXiv:0901.4352,priv:artoisenet:2009}. 
In order to avoid this problem, measurements and 
calculations were repeated increasing the $p_T$ cut first to 2 GeV and 
then to 3 GeV. In Fig.~\ref{fig:highpt}a) and b) the $\lambda$ and 
$\nu$ parameters, respectively, are shown as a function of $z$ for 
$p_T > 2$ GeV, while in Fig.~\ref{fig:highpt}c) and d) the same 
parameters are displayed for $p_T > 3$ GeV. All the values are listed 
in Table~\ref{tab:highpt}.
The NLO CS calculation~\cite{priv:artoisenet:2009}, also shown in these 
figures, has now smaller 
uncertainties, but the agreement with the data is only satisfactory 
for the $\nu$ parameter. Sizeable discrepancies remain for the $\lambda$ 
parameter both for $p_T > $~2\,~GeV and $p_T > $~3\,~GeV.

\section{Conclusions}
\label{sec:sum}

The $J/\psi$ helicity distributions in the inelastic photoproduction regime 
have been measured using a luminosity of 468~pb$^{-1}$. 
The $J/\psi$ helicity parameters $\lambda$ and $\nu$ were extracted  
in the target frame as a function of the transverse momentum and 
of the inelasticity of the $J/\psi$. 
The results were compared to LO QCD predictions in 
the colour-singlet, colour-singlet plus colour-octet and $k_T$ factorisation 
frameworks.
A recent NLO QCD prediction in the colour-singlet framework was also 
considered.
Even though the experimental and theoretical uncertainties 
are large, none of the predictions can describe all aspects of the data.

\section*{Acknowledgments}
\vspace{0.3cm}
We appreciate the contributions to the construction and maintenance 
of the ZEUS detector of many people who are not listed as authors.  
The HERA machine group and the DESY computing staff are especially 
acknowledged for their success in providing excellent operation of 
the collider and the data analysis environment.  
We thank the DESY directorate for their strong support and encouragement.
It is a pleasure to thank P.~Artoisonet, S.M.~Baranov, M.~Kr\"{a}mer 
and F.~Maltoni for helpful discussions and for providing their 
predictions.


{
\def\bibname{\Large\bf References}
\def\refname{\Large\bf References}
\pagestyle{plain}
\ifzeusbst
  \bibliographystyle{./BiBTeX/bst/l4z_default}
\fi
\ifzdrftbst
  \bibliographystyle{./BiBTeX/bst/l4z_draft}
\fi
\ifzbstepj
  \bibliographystyle{./BiBTeX/bst/l4z_epj}
\fi
\ifzbstnp
  \bibliographystyle{./BiBTeX/bst/l4z_np}
\fi
\ifzbstpl
  \bibliographystyle{./BiBTeX/bst/l4z_pl}
\fi
{\raggedright
\bibliography{./BiBTeX/user/syn.bib,%
              ./BiBTeX/bib/l4z_articles.bib,%
              ./BiBTeX/bib/l4z_books.bib,%
              ./BiBTeX/bib/l4z_conferences.bib,%
              ./BiBTeX/bib/l4z_h1.bib,%
              ./BiBTeX/bib/l4z_misc.bib,%
              ./BiBTeX/bib/l4z_old.bib,%
              ./BiBTeX/bib/l4z_preprints.bib,%
              ./BiBTeX/bib/l4z_replaced.bib,%
              ./BiBTeX/bib/l4z_temporary.bib,%
              ./BiBTeX/bib/l4z_zeus.bib,%
              ./BiBTeX/bib/l4z_my.bib}}
}
\vfill\eject

%
%
%
%
\begin{table}[p]
\begin{center}
\begin{tabular}{||c|c|c|c||}
\hline
$p_T$ range  (GeV) & $ \langle p_T \rangle $ (GeV)&  $\lambda$  & $\nu$ \\
\hline\hline
\(1.0\rnge 1.4\)   & 1.2  & \(~~0.16^{+0.15~+0.10}_{-0.16~-0.04}\)  
                          & \(~~0.26^{+0.10~+0.05}_{-0.10~-0.08}\)  \\
\(1.4 \rnge 1.9\)  & 1.6  & \(~~0.38^{+0.17~+0.10}_{-0.17~-0.10}\)  
                          & \(-0.12^{+0.12~+0.07}_{-0.14~-0.01}\)   \\
\(1.9 \rnge 2.4\)  & 2.1  & \(-0.15^{+0.17~+0.20}_{-0.17~-0.08}\)  
                          & \(-0.33^{+0.21~+0.12}_{-0.24~-0.16}\)   \\
\(2.4 \rnge 3.4\)  & 2.8  & \(~~0.21^{+0.17~+0.26}_{-0.16~-0.05}\)  
                          & \(-0.09^{+0.19~+0.13}_{-0.20~-0.13}\)   \\
\(3.4 \rnge 4.2\)  & 3.7  & \(~~0.34^{+0.32~+0.16}_{-0.28~-0.17}\)  
                          & \(~~0.21^{+0.26~+0.10}_{-0.28~-0.03}\)  \\
\(4.2 \rnge 10.\) & 5.2  & \(~~0.31^{+0.33~+0.21}_{-0.31~-0.19}\)  
                          & \(-0.50^{+0.27~+0.11}_{-0.28~-0.07}\)   \\
\hline 
\end{tabular}
\caption{
$J/\psi$ helicity parameters $\lambda$ and $\nu$ as a 
function of $p_T$ measured in the target frame for 50 $ < W < 180$ GeV, 
0.4 $ < z < 1$ and $p_T > 1$ GeV.
The first uncertainty is statistical and the second is systematic.}
\label{tab:lanuvspt}
\end{center}
\end{table}

\begin{table}[p]
\begin{center}
\begin{tabular}{||c|c|c|c||}
\hline
\(z\) range   & $ \langle z \rangle $  & $\lambda$ & $\nu$ \\ 
\hline \hline
\(0.10\rnge0.40\) & 0.27  & \(~~0.27^{+0.43~+0.13}_{-0.40~-0.14}\) 
                          & \(~~0.99^{+0.24~+0.14}_{-0.27~-0.40}\) \\
\(0.40\rnge0.55\) & 0.48  & \( -0.22^{+0.19~+0.18}_{-0.18~-0.06}\) 
                          & \(-0.01^{+0.17~+0.08}_{-0.18~-0.03}\) \\
\(0.55\rnge0.70\) & 0.61  & \(~~0.16^{+0.12~+0.10}_{-0.13~-0.04}\)
                          & \(-0.05^{+0.10~+0.05}_{-0.11~-0.02}\) \\
\(0.70\rnge0.80\) & 0.75  & \(~~0.39^{+0.15~+0.04}_{-0.15~-0.13}\) 
                          & \(-0.07^{+0.11~+0.04}_{-0.13~-0.07}\) \\
\(0.80\rnge0.90\) & 0.85  & \(~~0.30^{+0.17~+0.15}_{-0.17~-0.05}\) 
                          & \(~~0.04^{+0.14~+0.08}_{-0.15~-0.08}\) \\
\(0.90\rnge1.0~~ \) & 0.95  & \(~~0.49^{+0.39~+0.15}_{-0.34~-0.08}\) 
                          & \(~~0.54^{+0.26~+0.16}_{-0.28~-0.08}\) \\
\hline
\end{tabular}
\caption{
$J/\psi$ helicity parameters $\lambda$ and $\nu$ as a function of 
$z$ measured in the target frame for 50 $ < W < 180$ GeV,
0.1 $ < z < 1$ and $p_T > 1$ GeV. 
The first uncertainty is statistical and the second is systematic.}
\label{tab:lanuvsz}
\end{center}
\end{table}

\begin{table}[p]
\begin{center}
\begin{tabular}{||c|c|c|c|c||} 
\hline
$p_T$ range  (GeV) & \(z\) range & $\langle z \rangle$ & $\lambda$ & $\nu$ \\ 
\hline\hline
& \(0.10\rnge0.55\) & 0.37  & \(~~0.35^{+0.34~+0.04}_{-0.31~-0.41}\) 
                            & \(-0.37^{+0.35~+0.08}_{-0.39~-0.09}\) \\
$p_T > 2$ &\(0.55\rnge0.70\) & 0.61   & \(~~0.05^{+0.18~+0.20}_{-0.17~-0.06}\)
                            & \(-0.10^{+0.19~+0.08}_{-0.20~-0.05}\) \\
& \(0.70\rnge0.80\) & 0.75  & \(~~0.34^{+0.22~+0.04}_{-0.22~-0.17}\) 
                            & \(-0.05^{+0.21~+0.06}_{-0.23~-0.25}\) \\
& \(0.80\rnge0.90\) & 0.85  & \(~~0.12^{+0.24~+0.29}_{-0.23~-0.08}\) 
                            & \(-0.39^{+0.26~+0.22}_{-0.27~-0.01}\) \\
\hline\hline
& \(0.10\rnge0.55\) & 0.38  & \(~~0.80^{+0.53~+0.11}_{-0.45~-0.43}\) 
                            & \(~~0.07^{+0.40~+0.07}_{-0.44~-0.04}\) \\
$p_T > 3$ & \(0.55\rnge0.70\) & 0.62  & \(~~0.26^{+0.31~+0.23}_{-0.28~-0.06}\)
                            & \(-0.26^{+0.27~+0.20}_{-0.28~-0.06}\) \\
& \(0.70\rnge0.90\) & 0.79  & \(~~0.09^{+0.25~+0.27}_{-0.23~-0.12}\) 
                            & \(-0.35^{+0.27~+0.09}_{-0.28~-0.07}\) \\
\hline

\end{tabular} 
\caption{$J/\psi$ helicity parameters $\lambda$ and $\nu$ as a function 
of $z$ measured in the target frame for 50 $ < W < 180$ GeV, 0.1 $ < z < 1$ 
and for $p_T > 2$ and $p_T > 3$ GeV.
The first uncertainty is statistical and the second is
systematic.}
\label{tab:highpt}
\end{center}
\end{table}


%
%
\end{document}